\pdfoutput=1
\documentclass[a4paper,10pt,5p,preprint]{elsarticle}
\usepackage[utf8]{inputenc}

\usepackage{microtype}
\usepackage{amsmath,amsfonts,amssymb,amstext,bbm}
\usepackage{url}
\usepackage{fancyhdr}
\usepackage{latexsym}
\usepackage{epsfig}
\usepackage{graphicx}
\usepackage[textsize=footnotesize]{todonotes}

\usepackage{doi}
\usepackage{hyperref}
\hypersetup{
	colorlinks=true
}

\DeclareMathOperator{\Tr}{Tr}

\let\Re\relax
\DeclareMathOperator{\Re}{Re}
\let\Im\relax
\DeclareMathOperator{\Im}{Im}
\newcommand{\lambdaB}{\boldsymbol{\lambda}}

\journal{Physics Letters A}

\biboptions{sort&compress}

\begin{document}

\begin{frontmatter}

\title{A perspective on multiparameter quantum metrology: \\
from theoretical tools to applications in quantum imaging}


\author[uow,warwick]{F. Albarelli}
\author[romatre,cnr]{M. Barbieri}

\author[milano]{M.G. Genoni}

\cortext[mycorrespondingauthor]{Corresponding author}
\ead{marco.genoni@fisica.unimi.it}

\author[capienza,romatre]{I. Gianani}

\address[uow]{Faculty of Physics, University of Warsaw, 02-093 Warszawa, Poland}
\address[warwick]{Department of Physics, University of Warwick, Gibbet Hill Road, CV4 7AL, Coventry, United Kingdom}

\address[romatre]{Dipartimento di Scienze, Universit\`a degli Studi Roma Tre, 00146, Rome, Italy}

\address[cnr]{Istituto Nazionale di Ottica - CNR, 50125, Florence, Italy}

\address[milano]{Quantum Technology Lab, Dipartimento di Fisica ``Aldo Pontremoli'', Universit\`a degli Studi di Milano, 20133, Milan, Italy}

\address[capienza]{Dipartimento di Fisica, Sapienza Universit\`a di Roma, 00185, Rome, Italy}

\begin{abstract}
The interest in a system often resides in the interplay among different parameters governing its evolution. It is thus often required to access many of them at once for a complete description. Assessing how quantum enhancement in such multiparameter estimation can be achieved depends on understanding the many subtleties that come into play: establishing solid foundations is key to delivering future technology for this task. In this article we discuss the state of the art of quantum multiparameter estimation, with a particular emphasis on its theoretical tools, on application to imaging problems, and on the possible avenues towards the next developments.
\end{abstract}

\begin{keyword}
Quantum estimation, quantum sensing, quantum imaging, superresolution imaging 
\end{keyword}

\end{frontmatter}


\section{Introduction}
Quantum sensing took its inspiration from the potential application to delicate systems \cite{Giovannetti1330,Giovannetti2006,Giovannetti2011, PhysRevD.23.1693,PhysRevD.26.1817,Demkowicz-Dobrzanski2015a,Degen2016,BraunRMP2018, Berchera2018, Pirandola2018}.
Its theory proves to be effective in establishing under which conditions quantum systems may be preferable to classical resources to estimate one parameter. This is captured by the minimal uncertainty attainable, with a given amount of resources, as quantified by the quantum Cramér-Rao bound (CRB).
The extension to the multiparameter case is not as clear cut.
Indeed, the optimal measurement targeting one parameter might be at odds with the optimal scheme for a different one.
Incompatibility between the measurements in this case stems from fundamental relations in quantum mechanics.
In addition when the measurement is fixed, correlations may arise among the parameters.
All this translates in trade-off relations among the uncertainties on the parameters, since the initial information in the probe has now to be apportioned.
This makes the multiparameter optimization problem more involved and at the same time more intriguing, and consequently also the tools and methods at the basis of multiparameter estimation differ substantially from the simpler ones useful to investigate a single-parameter case~\cite{Szczykulska2016,Liu2019d}.
Failing to understand how these differences come about may prevent to attain a genuine advantage in a naive approach. It is then essential to inspect the multiparameter theory in all its features.   

Biological materials are the perfect example in that their complexity makes them susceptible to easily altering their properties and behaviours in the presence of nonlinear interactions~\cite{Mirmiranpour2018,Vojisavljevic2007,Pena2012}. 
In order to limit these, low illumination might be preferable, but requires optimizing its properties to mitigate potential loss due to limited signal-to-noise ratio. This is the goal of quantum metrology, which investigates the optimal state preparation and measurement to extract the maximum amount of information.
Archetypical systems in quantum metrology have been investigated focusing on a single parameter containing all relevant information~\cite{ae04,asp,pryde17,fabio13,Monras2007,Tischlere1601306,Yonezawa1514}.
The complexity of biological systems however, also reflects on the fact that many parameters at once, including possible parasitic processes, may be needed to capture even their essential features.

Imaging stands out as the most informative and direct technique for inspecting biological systems~\cite{Taylor2016, Cox}, with efforts aimed at circumventing limitations in the resolution by means of techniques including stimulated emission depletion microscopy (STED), stochastic optical reconstruction microscopy (STORM), and photoactivated localization microscopy (PALM) \cite{Hell07,reviewmicro,Betzig1642,rust06}.
The abundance of information it provides is granted by its intrinsically multiparameter approach. Therefore, advantages of quantum imaging, or even of classical imaging, are better understood in the multiparameter framework.

In this perspective article we discuss the multiparameter approach to quantum estimation, reviewing the main results that have led to the current understanding of the problem. We further discuss how potential developments may come from  imaging in either the superresolution or phase imaging schemes. 

\section{Theoretical tools for quantum multiparameter estimation: state of the art and perspective}
In this section we give a brief but comprehensive review of the main theoretical tools that are needed to address the problem of multiparameter estimation via quantum probes and quantum measurements, starting from the original formulations introduced by Helstrom~\cite{helstrom1976quantum} and Holevo~\cite{Holevo2011b}, along with the most recent fundamental results and open questions.
\subsection{Matrix quantum Cram\'er-Rao bounds}
A multiparameter quantum estimation problem is defined by a quantum statistical model $\varrho_{\boldsymbol \lambda}$, that is a family of density operator labelled by a vector of $d$ real unknown parameters $\lambdaB = (\lambda_1, \lambda_2, \dots, \lambda_d)^{\sf T}$.
In order to estimate the $d$ parameters one performs a quantum measurement described by a POVM $\Pi = \{ \Pi_k  \, | \, \Pi_k \geq 0, \, \sum_k \Pi_k = \mathbbm{1} \}$, yielding detection conditional probabilities through the Born rule, $p(k | \lambdaB ) = \Tr[ \varrho_{\lambdaB} \Pi_k ]$.
Assuming that the measurement is repeated $M$ times, obtaining a list of outcomes ${ \kappa} = \{ k_1, k_2, \dots, k_M\}$ (assumed to be independent and identically distributed), the parameters are then estimated through an estimator $\widetilde{\lambdaB}(\kappa)$, that is a map from the space of measurement outcomes to the space of the possible values of the parameters $\lambdaB \in \mathbbm{R}^d$, whose accuracy is typically addressed in terms of its mean-square error matrix
\begin{align}
{\bf V}(\lambdaB) = \sum_k p(\kappa |\lambdaB) \, \left(\widetilde\lambdaB(\kappa) - \lambdaB \right)  \left(\widetilde\lambdaB(\kappa) - \lambdaB \right)^{\sf T}\,.
\end{align}
By considering locally unbiased estimators that satisfy 
\begin{align}
\sum_k ( \widetilde{\lambda}_\mu(\kappa) - \lambda_\mu ) p(
\kappa|\lambdaB) = 0 \,, \,\,\,\,\, \sum_{k} \widetilde{\lambda}_\mu(\kappa) \frac{\partial p(\kappa|\lambdaB)}{\partial \lambda_\nu} = \delta_{\mu\nu} \,,
\end{align}
the classical CRB\footnote{Given two positive semidefinite matrices $\mathbf{A}$ and $\mathbf{B}$ we say that ${\bf A} \geq {\bf B}$ if ${\bf A} - {\bf B}$ is a positive semi-definite matrix; this defines a \emph{partial} ordering among such matrices (Loewner order). }
\begin{align}
{\bf V}(\lambdaB) \geq \frac{1}{M} {\bf F}(\lambdaB)^{-1} \,, \label{matrixCRB}
\end{align}
puts a constraint over all the possible mean-square error matrices in terms the classical Fisher information (FI) matrix with elements
\begin{align}
F_{\mu\nu} = \sum_k p(k|\lambdaB) \left( \partial_\mu \log p(k|\lambdaB) \right) \left( \partial_\nu \log p(k|\lambdaB) \right), \,\,\,
\end{align} 
where $\partial_\mu = \frac{\partial}{\partial \lambda_\mu}$, and holds for any fixed classical statistical model $p(k|\lambdaB)$ \cite{cramer1946mathematical}.
For locally unbiased estimators this bound is always attainable for any $M$.
With more realistic estimators the bound 
is attained in the limit of an infinite number of measurements $M$, e.g. by the maximum likelihood estimator, so that the mean-square error matrix is equal to the rescaled inverse of the FI matrix~\cite{lehmann_theory_1998}
\begin{align}
\lim_{M \to \infty} M \,{\bf V}(\lambdaB) &= {\bf F}(\lambdaB)^{-1}\,.
\end{align}
Because of this asymptotic attainability it is common, especially when studying quantum applications, to work with locally unbiased estimators~\cite{Holevo2011b,Suzuki2019a}, this amounts to assume that the the number of repetitions $M$ is high enough to guarantee attainability.
In the following we will do the same and drop the factor $M$ from our expressions.

In the quantum setting, that is when the conditional probability is obtained through the Born rule $p(k|\lambdaB) = \Tr[\varrho_{\lambdaB} \Pi_k ]$, one can obtain more general bounds that depend only on the quantum statistical model $\varrho_{\lambdaB}$ and not on the particular measurement strategy $\{\Pi_k\}$~\cite{Hayashi2005,Hayashi2017c}.
The most celebrated and useful approaches have been pursued by Helstrom~\cite{Helstrom1967} and Yuen and Lax~\cite{Yuen1973} (in parallel to Belavkin~\cite{Belavkin1976}) by introducing, respectively, the symmetric logarithmic derivative (SLD) operators $L_\mu^{\sf S}$, and the right logarithmic derivative (RLD)  operators $L_\mu^{\sf R}$, defined implicitly via the equations
\begin{align}
\partial_\mu \varrho_{\lambdaB} &= \frac{L_\mu^{\sf S}\varrho_{\lambdaB} + \varrho_{\lambdaB} L_\mu^{\sf S}}{2}, \label{eq:SLD}\\
\partial_\mu \varrho_{\lambdaB} &= \varrho_{\lambdaB} L_\mu^{\sf R} \,.
\end{align}
By means of the corresponding SLD $\mathbf{Q}(\lambdaB)$ and  RLD ${\bf J}(\lambdaB)$ quantum Fisher information (QFI) matrices, with elements  
\begin{align}
Q_{\mu\nu}(\lambdaB) &= \Tr \left[\varrho_{\lambdaB} \frac{ L_\mu^{\sf S}  L_\nu^{\sf S}  + L_\nu^{\sf S}  L_\mu^{\sf S}  }{2} \right], \\
J_{\mu\nu}(\lambdaB) &= \Tr \left[\varrho_{\lambdaB}  L_\nu^{\sf R} L_\mu^{\sf R\,\dag} \right] \,,
\end{align}
one can then derive the following (measurement independent) matrix quantum CRBs
\begin{align}
{\bf V}(\lambdaB) &\geq \mathbf{Q}(\lambdaB)^{-1}  & \,\,\, &\text{matrix SLD-CRB,} \label{eq:matrixQCRC^SLD} \\
{\bf V}(\lambdaB) &\geq {\bf J}(\lambdaB)^{-1}   &\,\,\, &\text{matrix RLD-CRB.}  \label{eq:matrixQCRC^RLD}
\end{align}
If a reparametrization is needed, i.e. if one wants to estimate a vector of parameters $\overline{\lambdaB}$, that is a function of the original parameters $\lambdaB$, the corresponding (classical and quantum) Fisher informations are obtained by following the same rules that hold for the mean-squared error matrix, i.e.
\begin{align}
\mathbf{Q}(\overline{\lambdaB}) = {\bf B} \,\mathbf{Q}(\lambdaB) \,{\bf B}^{\sf T},\,\,\,\,\, {\bf J}(\overline{\lambdaB}) = {\bf B} \,{\bf J}(\lambdaB) \,{\bf B}^{\sf T}  \,,
\end{align}
where the reparametrization matrix ${\bf B}$ is defined via its elements $B_{\mu\nu} = \partial \lambda_\nu / \partial\overline{\lambda}_\mu$ (i.e. the transpose of the Jacobian matrix).
Also the corresponding SLD and RLD operators can be easily obtained via the relationships
\begin{align}
\overline{L}_\mu^{\sf S} = \sum_\nu B_{\mu\nu} L_\nu^{\sf S} \,, \,\,\,\,\, \overline{L}_\mu^{\sf R} = \sum_\nu B_{\mu\nu} L_\nu^{\sf R}\,.
\end{align}

In the single-parameter scenario ($d=1$), the matrix CRBs (\ref{eq:matrixQCRC^SLD}) and (\ref{eq:matrixQCRC^RLD}) become scalar inequalities, and one can prove that the SLD bound is attainable since it always exists a POVM such that the corresponding classical FI is equal to the SLD-QFI~\cite{helstrom1976quantum,Nagaoka1989,Braunstein1994}.
In particular one proves that an optimal POVM, that attains the SLD-CRB, is the projection over the eigenstates of the corresponding SLD operator.

On the other hand in the multiparameter scenario the matrix CRBs (\ref{eq:matrixQCRC^SLD}) and \eqref{eq:matrixQCRC^RLD} are in general not attainable.
If we restrict to the SLD bound, its general non-attainability can be heuristically understood by considering parameters whose SLD operators, corresponding to optimal measurement strategies, do not commute.
Also for this reason, to get a better insight into the performance of different multiparameter estimators, it is customary to recast the matrix bounds into scalar bounds.
\subsection{Scalar quantum Cram\'er-Rao bounds}
Scalar CRBs are obtained by introducing a weight matrix $\mathbf{W}$ (positive, real matrix of dimension $d \times d$), such that one can derive the following scalar inequalities from Eqs.~\eqref{eq:matrixQCRC^SLD} and \eqref{eq:matrixQCRC^RLD}:
\begin{align}
\Tr[{\bf W\, V}] \geq  C^{\sf S}(\lambdaB, {\bf W})\,, \,\,\,\,\, \Tr[{\bf W\, V}] \geq  C^{\sf R}(\lambdaB, {\bf W})\,,
\end{align}
where the scalar SLD- and RLD-CRBs read\footnote{We denote with $\textrm{Re}({\bf A})$ and $\textrm{Im}({\bf A})$ respectively the real and imaginary part of a complex-valued matrix ${\bf A}$.}
\begin{align}
C^{\sf S}(\lambdaB, {\bf W}) &= \Tr[ {\bf W}\, \mathbf{Q}^{-1}] \,, \\
C^{\sf R}(\lambdaB, {\bf W}) &= \Tr[ {\bf W}\, \textrm{Re}({\bf J}^{-1}) ] + \Vert \sqrt{ {\bf W}}\, \mathrm{Im}(\mathbf{J}^{-1}) \sqrt{\mathbf{W}} \Vert_1 \,,
\end{align} 
where we have introduced the trace norm $\Vert \mathbf{A} \Vert_1 = \Tr [ \sqrt{\mathbf{A}^{\dag} \, \mathbf{A}} ]$
For example, by considering ${\bf W} = \mathbbm{1}_d$, i.e. the identity matrix of dimension $d$, the inequalities above bound the quantity $\Tr[{\bf V}]$, i.e. the sum of the mean square errors for each unknown parameter $\lambda_\mu$.

Like the corresponding matrix bounds, these scalar bounds are in general not attainable.
We can define the \emph{most informative bound}, a minimization of the classical scalar bound over all the possible POVMs,
\begin{align}
\label{eq:MIbound}
C^{\sf MI}(\lambdaB, {\bf W}) &= \min_{\tiny \textrm{POVM}\,\, \Pi} \left[\Tr[{\bf W}\, {\bf F}^{-1} ] \right]
\end{align}
which is in general larger than the maximum of the two quantum CRBs.
At this point it is fundamental to remark that, while such a quantity will be attainable by performing quantum measurements on a single copy at a time, in general this procedure will be adaptive~\cite{Nagaoka1989a,Hayashi1997a,Gill2000,Fujiwara2006}.
This is true already at the single parameter level~\cite{Barndorff-Nielsen2000} and it is due to the fact that the optimal POVM in~\eqref{eq:MIbound} depends in general on the true value of the parameter.
In other words, we are implicitly assuming to have already enough information to implement a good approximation to the optimal measurement. This approach is commonly called \emph{local} quantum estimation theory~\cite{Paris2009}.

A tighter bound $C^{\sf H}(\lambdaB,{\bf W})$ has been derived by Holevo~\cite{Holevo1977,Holevo2011b}, such that the following chain of inequalities holds
\begin{align}
\Tr[{\bf W\, V}] \geq  C^{\sf MI}(\lambdaB, {\bf W}) \geq C^\mathsf{H}(\lambdaB, {\bf W}) \\
\geq \textrm{max}\left[ C^{\sf S}(\lambdaB, {\bf W}), C^{\sf R}(\lambdaB, {\bf W})\right] \,, \nonumber
\end{align}
The Holevo-CRB $C^{\mathsf{H}}(\lambdaB, {\bf W})$ is defined via the following minimization
\begin{align}
C^{\sf H}(\lambdaB, {\bf W}) =  \min_{{\bf U} \in \mathbbm{S}^d, {\bf X} \in \mathbbm{X}_{\lambdaB} }  \left[ \Tr[ {\bf W}\,{\bf U} ] \,\, | \,\,{\bf U} \geq {\bf Z}[{\bf X}] \right] \\
= \min_{{\bf X} \in \mathbbm{X}_{\lambdaB} }  \left[ \Tr[ {\bf W} \,  \Re {\bf Z}[{\bf X}] ] + \Vert \sqrt{\mathbf{ W}} \,  \Im {\bf Z}[{\bf X}] \sqrt{\mathbf{W}} \Vert_1  \right] \,, \label{eq:HolevoBound}
\end{align}
where $\mathbbm{S}^d$ denotes the set of real symmetric $d$-dimensional matrices, and the Hermitian $d \times d$ matrix ${\bf Z}$ is defined via its elements
\begin{align}
Z_{\mu \nu} \left[{\bf X}\right]= \Tr[ \varrho_{\lambdaB} X_\mu X_\nu ] \,
\end{align}
with the collection of operators ${\bf X}$ belonging to the set
\begin{align}
\mathbbm{X}_{\lambdaB} = \left\{ {\bf X} = (X_1, \dots , X_d )\,\, |\,\,  \Tr[( \partial_\mu \varrho_{\lambdaB} ) X_\nu ]= \delta_{\mu\nu} \right\} \,.
\end{align}

We remark that in general $C^{\sf H}$ is smaller than the most informative bound $C^\mathsf{MI}$ and it is possible to derive other intermediate bounds, as for example shown in~\cite{Nagaoka1989} for the two-parameter problem ($d=2$).
However the Holevo-CRB $C^\mathsf{H}$ is typically regarded the most fundamental scalar bound for multiparameter quantum estimation, as it is proven to be equivalent to the most informative bound of the asymptotic model, i.e. it becomes attainable by performing a collective measurement on an asympotically large number of copies of the state $\varrho_{\lambdaB}^{\otimes n} = \bigotimes_{j=1}^n \varrho_{\lambdaB}$, with $n \to \infty$~\cite{Hayashi2008a,Kahn2009,Yamagata2013,Yang2018a}.
There are however instances where it is proven that the standard Holevo-CRB is in fact equivalent to the most-informative bound in the single-copy scenario, and thus attainable via separable measurements: this is the case of pure state states~\cite{Matsumoto2002} and displacement estimation with Gaussian states~\cite{Holevo2011b}.

The minimization that one has to perform in Eq.~\eqref{eq:HolevoBound} makes the evaluation of $C^{\mathsf{H}}$ not straightforward and closed form expressions for non-trivial cases are hard to obtain.
To the best of our knowledge all known analytical results for cases in which the Holevo-CRB does not trivially reduce to the SLD- or RLD-CRBs (more details on this in Sec.~\ref{sec:classSuzuki}) have only been obtained for two-parameter estimation problems.
There are generic results for qubit systems~\cite{Suzuki2016a} and for two-parameter estimation with pure states~\cite{Matsumoto2002} and also particular results for two-parameter displacement estimation with two-mode Gaussian states in~\cite{Bradshaw2017a,Bradshaw2017}, where, thanks to the calculation of the Holevo-CRB, it was possible to complete the partial results that were previously obtained in~\cite{Genoni2013b} via the calculations of the SLD and RLD bounds only.
A general closed-form expression for two parameter problems was given recently in~\cite{Sidhu2019a}.

In~\cite{Albarelli2019}, it was shown that the minimization in~\eqref{eq:HolevoBound} is a convex problem and that the evaluation the Holevo-CRB for finite-dimensional systems can be recast as a semidefinite program.
Numerical results for relevant multi-parameter estimation problems, such as estimation of phase and loss in interferometric schemes~\cite{Crowley2014} and 3D-magnetometry~\cite{Baumgratz2015}, have been presented, providing also numerical evidence of other regimes and quantum estimation problems where the bound becomes attainable with single-copy measurements.
The Holevo-CRB has been investigated numerically also in the context of error-corrected multiparameter quantum metrology~\cite{Gorecki2019}.

More recently it was proved that $C^{\sf H}$ can be also upper bounded, as it satisfies the inequalities~\cite{Carollo2019,Albarelli2019b} (see also~\cite{Tsang2019c} for a similar result)
\begin{align}
C^{\sf S} (\lambdaB , {\bf W} ) &\leq C^{\sf H}(\lambdaB  , {\bf W}) \\
&\leq C^{\mathsf{S}}(\lambdaB , {\bf W} ) + \Vert \sqrt{\mathbf{W}}  \, \mathbf{Q}^{-1} \, \mathbf{D} \, \mathbf{Q}^{-1}\, \sqrt{\mathbf{W}} \Vert_1  \\
 &\leq ( 1 + \mathcal{R})\, C^{\sf S} (\lambdaB , {\bf W} )  \leq 2 \, C^{\sf S} (\lambdaB , {\bf W} )\,,
\label{eq:carollo}
\end{align}
where we have introduced the (asymptotic) \emph{incompatibility} matrix ${\bf D}$, also known as mean Uhlmann curvature~\cite{Carollo2018a},
with elements
\begin{align}
D_{\mu\nu} = -\frac{i}{2} \Tr[ \varrho_{\lambdaB} [L_\mu^{\sf S}, L_\nu^{\sf S}] ] \, ,
\end{align}
and the quantity
\begin{align}
\mathcal{R} = \lvert\vert i \, \mathbf{Q}^{-1} {\bf D} \rvert\rvert_\infty \, , \label{eq:Rdefinition}
\end{align}
where $\lvert\lvert {\bf A} \rvert\rvert_\infty$ denotes the largest eigenvalue of the matrix ${\bf A}$.
One can prove that $0\leq \mathcal{R} \leq 1$~\cite{Carollo2019}, or analogously that 
$ \Vert \sqrt{\mathbf{W}} \, \mathbf{Q}^{-1} \, \mathbf{D} \, \mathbf{Q}^{-1} \sqrt{\mathbf{W}} \Vert_{1} \leq C^{\mathsf{S}}(\lambdaB , {\bf W} )$~\cite{Albarelli2019b} and thus the scalar SLD-CRB gives in fact an estimate of the Holevo bound up to a factor $2$. 
Consequently, the Holevo-CRB cannot provide new information about possible quantum enhancements in scaling that is not already available in the SLD-CRB.
However, it can still provide novel insights into quantum sensing and estimation impossible via the SLD-CRB, for instance in simultaneous estimation of phase and loss in optical interferometry~\cite{Albarelli2019}.
Furthermore, the parameter $\mathcal{R}$ has been introduced as a measure of (asymptotic) incompatibility, as it will be more clear in the following.

%
An alternative scalar figure of merit has also been commonly considered to asses the performance of a given measurement scheme.
Given a quantum statistical model $\varrho_{\lambdaB}$, one can introduce the following quantity~\cite{Gill2000}
\begin{align}
\Upsilon(\varrho_{\lambdaB},\Pi) &= \Tr[ {\bf F}(\lambdaB)\, \mathbf{Q}(\lambdaB)^{-1}] \,, \nonumber \\
&= \sum_{\mu=1}^{d} \frac{F_{\mu\mu}(\lambdaB)}{Q_{\mu\mu}(\lambdaB)}\,\,\,\,\,\,\, \textrm{for diagonal}\,\, \mathbf{Q}(\lambdaB) \, \label{eq:zeta}
\end{align}
where ${\bf F}(\lambdaB)$ denotes the classical Fisher information corresponding to the considered measurement $\Pi$. 
It has been proven~\cite{Gill2000} that

\begin{align}
0\leq \Upsilon(\varrho_{\lambdaB},\Pi) \leq \min\left[d, \, \textrm{dim}(\mathbbm{H})-1 \right] \,,  \label{eq:zetabound}
\end{align} 
where we remind that $d$ is the number of parameters to be estimated in the vector $\lambdaB$, while $\textrm{dim}(\mathbbm{H})$ denotes the dimension of the Hilbert space of the quantum statistical model $\varrho_{\lambdaB}$.
Clearly the upper bound $d$ is achieved whenever all the parameters $\lambdaB$ can be simultaneously estimated at the ultimate quantum limit dictated by the SLD-QFI matrix. On the other hand, this inequality proves how the dimension of the Hilbert space imposes a trade-off on the ultimately achievable estimation precision for the whole set of parameters; for example if one restricts to qubits one obtains that $0\leq \Upsilon(\varrho_{\lambdaB},\Pi) \leq 1$.
This problem has been extensively studied both in the context of estimation of unitary operations\cite{Ballester2004,Ballester2004a} and of estimation of phase and phase-diffusion~\cite{Vidrighin2014,Altorio2015a,Roccia2017}.
It is important to remark that in cases where the econding of the parameters $\mathcal{E}_{\lambdaB}$ is fixed and one can optimize over the possible input probe states $\varrho_0$, such that $\varrho_{\lambdaB}=\mathcal{E}_{\lambdaB}(\varrho_0)$, the results of the optimization for different figures of merit such as $\Upsilon(\varrho_{\lambdaB},\Pi)$ or $\Tr[{\bf W\, V}]$ will in general give different results: different probe states will in fact yield different SLD-QFI matrices $\mathbf{Q}(\lambdaB)$, and while $\Upsilon(\varrho_{\lambdaB},\Pi)$ maximizes the joint optimal estimability at fixed $\mathbf{Q}(\lambdaB)$, the quantity $\Tr[{\bf W\, V}]$ maximizes the overall (weighted) precision for the parameters $\lambdaB$. An example of this kind has been studied in \cite{Roccia2017a}.

\begin{figure}[h]
\includegraphics[width=\columnwidth]{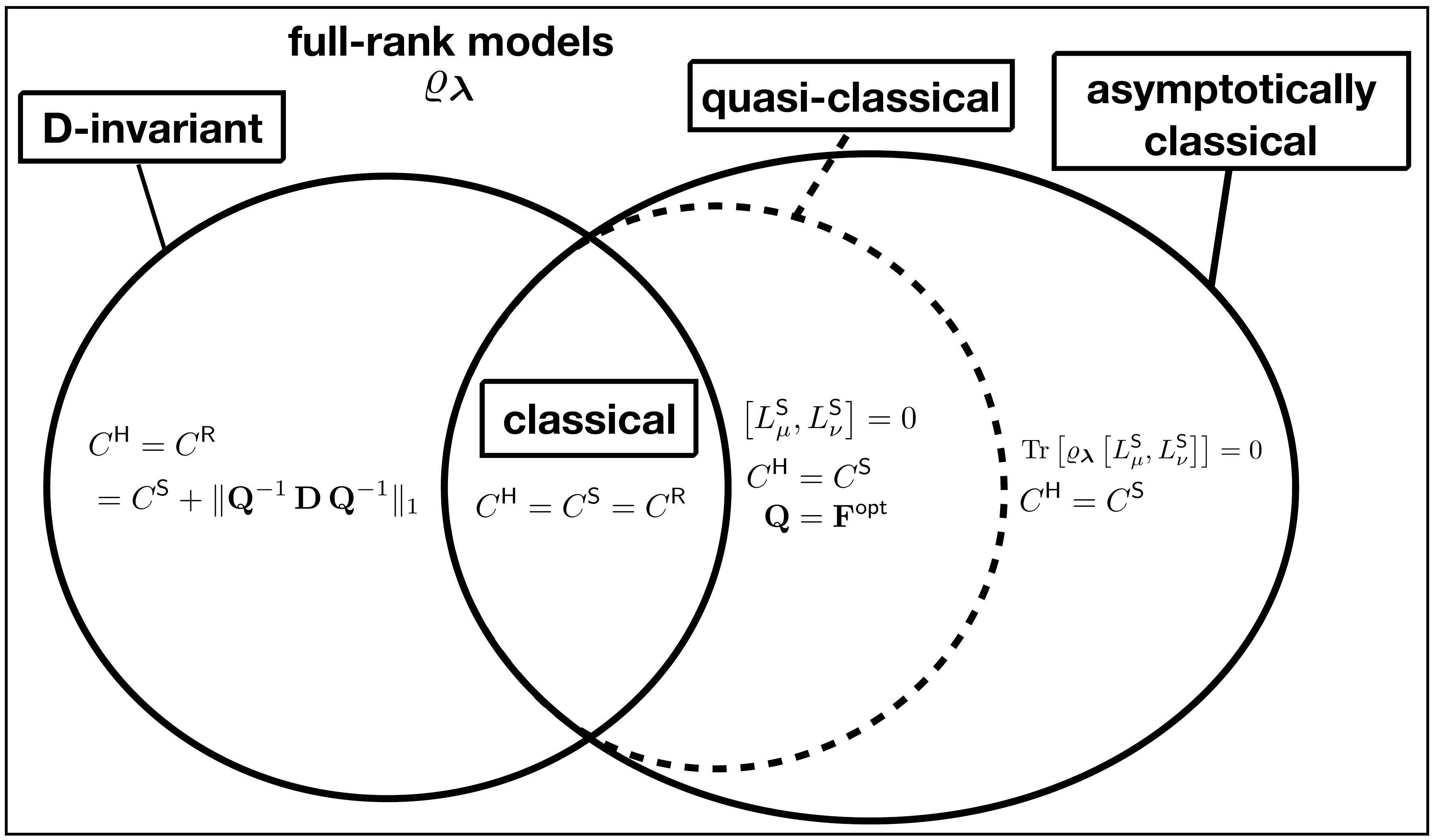}
\includegraphics[width=\columnwidth]{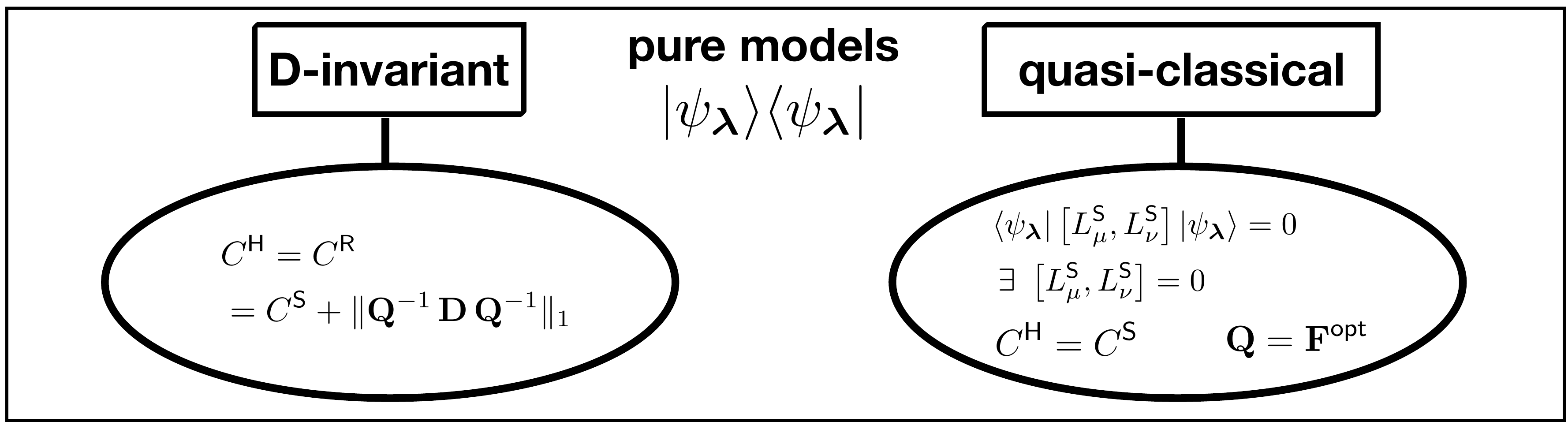}
\caption{Pictorial representation of the classification of quantum statistical models originally introduced in~\cite{Suzuki2018} and reviewed in Sec.~\ref{sec:classSuzuki}.
Top panel: the classification for full-rank models~\cite{Suzuki2018}, see discussion in Sec.~\ref{s:theopersp} for rank-deficient models.
Classical models are the intersection of D-invariant and asymptotically classical models.
Quasi-classical models are a proper subset of asymptotically classical ones, for which the SLDs commute and thus an optimal POVM whose FI matrix $\mathbf{F}^\mathsf{opt}$ attains the SLD-QFI matrix exists.
Bottom panel: classification for pure-state (rank-1) models.
The sets of asymptotically classical and quasi-classical coincide in this case, since it is always possible to find a set of commuting SLDs if the weak commutativity condition~\eqref{eq:weakcomm} is satisfied. There is no intersection between quasi-classical and D-invariant models (also called coherent in the pure case~\cite{Fujiwara1999}).
In both panels the $C^\mathsf{R}$ bound in the D-invariant case is reported for ${\bf W}=\mathbbm{1}_d$ for brevity.
}
\label{f:Suzukiclasses}
\end{figure}
\subsection{A classification of multiparameter quantum statistical models}\label{sec:classSuzuki}
By studying in more detail the relationship between the different bounds, it is possible to classify multiparameter quantum estimation problems, as shown by Suzuki in~\cite{Suzuki2018} and pictured in Fig.~\ref{f:Suzukiclasses}.
We can identify the following classes:
\begin{itemize}
\item {\bf Classical} quantum statistical models

A quantum statistical model is said classical if $\varrho_{\lambdaB}$ can be diagonalized with a $\lambdaB$-independent unitary, i.e. $\varrho_{\lambdaB} = U \Lambda_{\lambdaB} U^{\dag}$, with $\Lambda_{\lambdaB}$ being a diagonal operator.
These models are dubbed classical, as they can be described in terms of a completely classical statistical model, where the quantum density operators are replaced by classical probability distributions. In this instance the SLD-CRBs (matrix and scalar) are naively attainable, by considering as the optimal measurement for all the parameters the diagonal basis of $\varrho_{\lambdaB}$, yielding a classical Fisher information satisfying ${\bf F}^\mathsf{opt}(\lambdaB) = \mathbf{Q}(\lambdaB)=\mathbf{J}(\lambdaB)$ and thus $C^{\mathsf{S}}(\lambdaB , {\bf W} )= C^{\mathsf{R}}(\lambdaB , {\bf W} ) =C^{\mathsf{H}}(\lambdaB , {\bf W} ) \,\, \forall \,\, W $.
\item {\bf Quasi-classical} quantum statistical models

A quantum statistical model $\varrho_{\lambdaB}$ is said quasi-classical if all SLD operators commute with each other at all points $\lambdaB$, that is
\begin{align}
[ L_\mu^{\sf S}, L_\nu^{\sf S} ] = 0\,, \,\,\, \forall \mu, \nu,\,\textrm{and}\,\, \forall \lambdaB \,.
\end{align}
While in general it is not possible to rephrase the estimation problem as a classical one, these models are dubbed {\em quasi-classical} as, since the SLD operators commute, it is possible to perform a measurement that saturates the SLD-CRB exactly, i.e. yielding also in this case a classical Fisher information ${\bf F}^\mathsf{opt}(\lambdaB) = \mathbf{Q}(\lambdaB)$, already at the single-copy level.
It then follows that $C^{\mathsf{S}}(\lambdaB , {\bf W} ) = C^{\mathsf{H}}(\lambdaB , {\bf W} ) \, \forall \, W $.

\item {\bf D-invariant} quantum statistical models

A quantum statistical model $\varrho_{\lambdaB}$ is called D-invariant at $\lambdaB$, it the SLD-tangent space at $\lambdaB$ is an invariant subspace of the commutation superoperator.
Mathematically,
\begin{align}
\forall X \in \mathcal{T}(\varrho_{\lambdaB}) \rightarrow \mathcal{D}_{\varrho_{\lambdaB}}(X) \in \mathcal{T}(\varrho_{\lambdaB})  \,
\end{align}
where we have defined the SLD tangent space, as the linear span of SLD operators,
\begin{align}
\mathcal{T}(\varrho_{\lambdaB}) = \textrm{span}_{\mathbbm{R}} \{ L_\mu^{\sf S} \} \,.
 \end{align}
and the commutation superoperator $\mathcal{D}_{\varrho_{\lambdaB}}$ via the equation
\begin{align}
\varrho_{\lambdaB} X - X \varrho_{\lambdaB} = i \left( \varrho_{\lambdaB} \mathcal{D}_{\varrho_{\lambdaB}}(X) + \mathcal{D}_{\varrho_{\lambdaB}}(X) \varrho_{\lambdaB}\right)  \,.
\end{align}
Remarkably, if this condition is fulfilled the Holevo-CRB coincides with the scalar RLD-CRB and also with the first upper bound in~\eqref{eq:carollo}, as proven in~\cite{Suzuki2016a}:
\begin{equation}
\begin{split}
&C^\mathsf{R}(\mathbf{W})=C^{\sf H}(\mathbf{W})\\
&=C^{\mathsf{S}}({\bf W} ) + \Vert \sqrt{\mathbf{W}}  \, \mathbf{Q}^{-1} \, \mathbf{D} \, \mathbf{Q}^{-1} \, \sqrt{\mathbf{W}} \Vert_1 \,\,\,\,  \forall \,{\bf W}\,,
\end{split}
\end{equation}
indicating that the RLD bound can be attained by a collective measurement on an asymptotically large number of copies $\varrho_{\lambdaB}^{\otimes n}$.
An important class of D-invariant models arises in finite-dimensional quantum state tomography, i.e. estimating all the parameters in a density matrix~\cite{Huangjun2012}.
\item \textbf{Asymptotically classical} quantum statistical models

A quantum statistical model $\varrho_{\lambdaB}$ is called asymptotically classical if all the SLD-operators commute on average on $\varrho_{\lambdaB}$, a property also called \emph{weak commutativity}, that is
\begin{align}\label{eq:weakcomm}
D_{\mu\nu} = \frac{1}{2 i} \Tr[ \varrho_{\lambdaB} [L_\mu^{\sf S}, L_\nu^{\sf S}] ] = 0 \,, \,\,\,\,  \forall \mu, \nu,\,\textrm{and}\,\, \forall \lambdaB \,.
\end{align}
One proves that if and only if this condition is fulfilled, then the Holevo-CRB is equal to the the SLD-CRB~\cite{Ragy2016},
\begin{align}
C^{\sf H}(\lambdaB, {\bf W}) = C^{\sf S}(\lambdaB, {\bf W}) \,\,\,\,  \forall\, {\bf W}\,.
\end{align}
In this instance, one has then that the SLD-CRB can be saturated asymptotically by considering a collective measurement on an asymptotically large number of copies $\varrho_{\lambdaB}^{\otimes n}$.

Given the definition above of asymptotically classical models, and the chain of inequalities (\ref{eq:carollo}), the quantity $\mathcal{R}$ defined in Eq. (\ref{eq:Rdefinition}) has been introduced in~\cite{Carollo2019} as a measure of {\em quantumness} of the quantum statistical model, or in other words of (asymptotic) incompatibility.
Indeed, it is zero if and only if the incompatibility matrix ${\bf D}$ is equal to the zero matrix, and thus if and only if the quantum statistical model is asymptotically classical.

In order to better understand the properties of certain asymptotically classical models, the behaviour of the quantity $\Upsilon(\varrho_{\lambdaB}^{\otimes n},\Pi_n)$ was studied in detail, where $\Pi_n$ denotes a, possibly collective, measurement acting on $n$-copies of the original state $\varrho_{\lambdaB}$.
In particular it was indeed shown, both by numerical simulations, theoretical calculations and experimental validation, that for a specific asymptotically classical model (estimation of phase and phase-diffusion), collective measurements on multiple copies of the qubit system described by $\varrho_{\lambdaB}$ allow to beat the single-copy qubit bound (\ref{eq:zetabound}), leading to the conjecture that the maximum value of $\Upsilon(\varrho_{\lambdaB}^{\otimes n},\Pi)$ increases monotonically with $n$, towards the asymptotic maximum value $\Upsilon(\varrho_{\lambdaB}^{\otimes n},\Pi_n) = d$, for $n \to \infty$.

\end{itemize}

\subsection{Discussion and perspective}\label{s:theopersp}
\noindent
The theory of multiparameter quantum estimation is an extremely challenging field, as witnessed by the fact that investigations on the topic started in the late 1960s~\cite{Helstrom1967,helstrom1976quantum}, but new theoretical results are still being found, e.g.~\cite{Pezze2017,Yang2018a,Suzuki2018,Albarelli2019}.
This is related to the fact that there are still many open questions in the field and here we highlight some of them.

Most of the literature on quantum parameter estimation has focused on $\mathrm{Tr}[\mathbf{W} \, \mathbf{F}(\lambdaB)^{-1} ]$ as the figure of merit.
The need to pick a scalar figure of merit arises because the ordering among positive semidefinite matrices is only partial.
Therefore it is not possible in general to optimize with respect to matrix inequalities.
This situation arises already at the classical level in the field of optimal design of experiments, where other scalar figures of merit are commonly considered too~\cite{Pukelsheim2006}.
This approach is starting to be applied also to quantum estimation~\cite{Gazit2019a,Lu2019} and we expect to see further progress in adapting ideas from the classical to the quantum domain.

However, there are some instances when it is possible to find POVMs that give an optimal \emph{matrix} bound.
This is the case when the FI matrix of a POVM exactly attains the SLD-QFI matrix.
Necessary and sufficient conditions for a POVM to satisfy the matrix equality between classical FI and QFI have been derived recently~\cite{Pezze2017,Yang2018a}.
Interestingly, for pure states such a POVM attaining the equality exists if an only if the weak commutativity conditions are satisfied; in turn, this is equivalent to the existence of commuting SLDs~\cite{Matsumoto2002}.
This means that asymptotically classical models and quasiclassical models coincide for pure states (see the bottom panel of  Figure~\ref{f:Suzukiclasses}).

While the conditions for the attainability of the matrix SLD-CRB for rank-1 (pure) and full-rank models are fully known, there is still work to do for arbitrary (but fixed) rank models\footnote{The case of \emph{variable} rank models is somewhat pathological, being akin to non-regular models in classical statistics~\cite{Seveso2019}.}.
When the quantum states in the model are not full rank the defining equation for the SLDs~\eqref{eq:SLD} does not specify those operators completely and they have some arbitrary components, which however do not affect the SLD-QFI matrix~\cite{Holevo2011b,Fujiwara1995,Liu2014a,Liu2014c}.
Nonetheless, these arbitrary component can affect other properties of the SLD operators, such as commutativity, allowing, as we pointed out above, to find commuting SLD-operators whenever the weak-commutativity condition is satisfied for pure states.
For arbitrary rank quantum statistical models, a necessary condition for finding a POVM with a FI matrix that attains the SLD-QFI is that the SLDs must commute on the support of the state (i.e. the orthogonal susbpace to the kernel of the density matrix), this has been dubbed \emph{partial commutativity}~\cite{Yang2018b} of the SLDs.
Whether or not this condition is also sufficient or equivalent to the existence of some commuting SLDs (by appropriately choosing the unspecified components) is still an important open problem.

One more optimization problem appears when the encoding of the parameters is fixed, but it is possible to optimize over the initial probe state.
Finding optimal states for quantum metrology is non-trivial already at the single parameter level~\cite{Frowis2014,Yuan2017a}.
Well-performing classes of states have been found for interesting multiparameter quantum metrology applications~\cite{Humphreys2013,Baumgratz2015}, but the problem has yet to be explored in detail, especially for simultaneous estimation of parameters with non-commuting generators.
In such a scenario it is often useful to use ancillary systems to overcome the incompatibility of the quantum statistical model, see e.g.~\cite{Fujiwara2001a,Imai2007,Chen2017a,Gorecki2019}.

Finally, we also mention that in various scenarios one might be interested only in the estimation of a subset of the whole vector of unknown parameters, while the remaining ones can be considered as \emph{nuisance parameters}.
Dealing with estimation in the presence of nuisance parameters is a well-known topic in classical estimation theory, which is only starting to be explored in the quantum domain~\cite{Suzuki2014,Yang2018a,Suzuki2019,Suzuki2019a}.
Particular applications of this framework are, for instance, distributed quantum sensing, where the aim is to estimate a linear combination (or more generally a function) of phases encoded by spatially separated generators~\cite{Proctor2018,Ge2017,Eldredge2018,Qian2019,Oh2019d,Sekatski2019}, and gradient magnetometry, where in order to estimate the gradient of the magnetic field, the homogeneous field acting on the atomic ensemble can be in fact considered as a nuisance parameter~\cite{Apellaniz2018}.
Furthermore, when the set of nuisance parameter is infinite-dimensional, more sophisticated tools are needed; in classical statistics this is the field of semiparametric estimation.
This approach has been applied to incoherent imaging by Tsang~\cite{Tsang2019b}, who has also started to generelize these ideas to quantum estimation~\cite{Tsang2019}.

\section{Applications to imaging}

Obtaining some form of advantage using quantum light has a history going back at least two decades. The interest has focused at first on ghost imaging~\cite{PhysRevLett.74.3600, Pittman1995, PhysRevA.57.3123, doi:10.1098/rsta.2016.0233} - the possibility of reconstructing the image of an object, hit by a reference beam, by analysing light on a second correlated beam, based on cross-correlation. Even though this phenomenon can be partially replicated with thermal light~\cite{PhysRevLett.89.113601, PhysRevLett.94.063601}, the use of quantum light grants some genuine enhancement, depending on the working conditions~\cite{PhysRevLett.92.033601,PhysRevA.82.053803}. The applications of correlations now stretch  far beyond their intended use for ghost imaging, also in the frequency domain~\cite{PhysRevA.78.061802, PhysRevLett.104.253603, PhysRevLett.113.160401, Ribeiro:97, Brida:2010fj,Kalashnikov:2016uq, Lemos:2014kx, PhysRevLett.119.243602,Losero2019arXiv,howard19}.
Quantum engineering of phase-sensitive states \cite{dowling02,Afek879,mitchell13,ian10,ian11,jeremy11,Nagata726,ae04,asp,pryde11,pryde17,ul15,fabio13,nico12} has also been suggested as a way to realise quantum-enhanced two-photon lithography~\cite{PhysRevLett.85.2733, PhysRevLett.87.013602, PhysRevA.77.012324} and microscopy~\cite{Ono2013}. Intensity squeezing is also being investigated to improve the signal to noise ratio in imaging, by lowering the noise in photon detection below the shot noise limit~\cite{PhysRevLett.93.243601, PhysRevLett.59.2555, PhysRevLett.88.203601, PhysRevLett.101.233604, PhysRevLett.102.213602, Treps940, Boyer544, Dowran:18, Taylor:2013fk, PhysRevX.4.011017}. 

Some of these applications are intimately connected to parameter estimation, when a single element of interest, for instance the position of a scattering centre, can be isolated~\cite{Taylor:2013fk, Tsang:15}. These cases are extended directly to the multiparameter realm when several such quantities are relevant to the probing. A different instance considers phase imaging, the reconstruction of an object by means of the optical phase acquired upon transmission.
This naturally leads to a description in terms of multiple phase estimation. This section is devoted to covering these two important approaches.
The following applications lay their ground on the concepts described in the previous section, however we remark how the potential of the multiparameter estimation theoretical framework has only been slightly scraped by the state of the art here reported.
We anticipate that taking fully advantage of the sophisticated mathematical tools previously discussed will lead to substantial advances in these applications and to the understanding of their fundamental limitations.

\subsection{Multiparameter estimation in separation measurements}

The theory of precise measurements of a small transverse separation of two incoherent point sources has been exhaustively covered in~\cite{Tsang2016b, Nair:16, PhysRevLett.117.190801, PhysRevLett.117.190802}, and generalized in terms of quantum superresolution~\cite{Gefen2019}. This two-source model is useful for understanding many subtleties, and is actually relevant for some applications - for instance, in astronomical observations~\cite{Tsang2019b} and imaging of luminescent targets~\cite{Willig2006, JPWolf}.
Though the description is conveniently cast in the quantum formalism, considering single photons, the treatment may actually apply to classical fields as well.

The individual positions of the sources, $X_1$ and $X_2$ in the object plane, are extracted by means of an imaging system with point spread function (PSF) $\psi(x)$, where $x$ is the coordinate in the image plane~\cite{Tsang2018}. For convenience, the PSF is taken to be spatially invariant and real, conditions that can be met in the experiment, at least on a limited portion of the available range. The standard direct approach would measure the photon flux $I(x)$ at every point $x$. For two sources with the exact same intensity, this has the form $I(x)=I_0(|\psi_1(x)|^2+|\psi_2(x)|^2)$, where $\psi_s(x)$ is the PSF associated to the source $i$, and $I_0$ is the total flux coming from the sources. The parameters of interest in this problem are the centroid of the source positions $\lambda_1=(X_1+X_2)/2$, and their difference $\lambda_2=(X_1-X_2)$. Calculation of the classical Fisher information matrix $\bold{F} (\lambdaB)$ reveals that this is diagonal, and that the $F_{22}$ term vanishes as $\lambda_2 \rightarrow 0$. This implies that arbitrary small separations can not be estimated, a fact called Rayleigh's curse~\cite{Tsang2016b, Paur:18}: the ghost of a late 1800's semi-empirical rule keeps haunting sophisticated modern-day imaging systems!

The choice of direct flux measurement is not unique: the possibility of devising better strategies should be ascertained by means of the SLD-QFI matrix $\bold{Q}(\lambdaB)$.
For PSFs of the kind we are considering, the matrix is diagonal, and, remarkably, its element $Q_{22}$ is a constant~\cite{Tsang2016b}; in particular, for Gaussian PSFs with width $\sigma$, the smallest achievable variance on $\lambda_2$ is $V_{22}(\lambdaB) = 4\sigma^2/M$. This can be made arbitrarily small by collecting a sufficiently large number of events, or, equivalently for classical light, sufficiently large intensity.    

While Rayleigh's curse can in principle be defeated, one needs to find an actual measurement achieving this condition. The authors of~\cite{Tsang2016b, Tsang_2017, PhysRevA.97.023830} introduce spatial demultiplexing on the Hermite-Gauss modes (SPADE) as a convenient choice in the presence of Gaussian PSFs. The associated classical Fisher information attains the ultimate quantum value, for fixed known centroid position.

Since the introduction of this method, a number of experiments have appeared presenting proof-of-principle implementations of this imaging technique. The authors of~\cite{Paur:18} looked at both Gaussian and sinc PSFs. The two point sources are created with a digital micromirror chip illuminated with a He-Ne laser, and shaped by means of circular (for the Gaussian) and square (for the sinc) apertures. The projective measurement is then performed by means of a spatial light modulator (SLM) imposing a hologram. The diffraction pattern was then imaged by a lens, ensuring separation of the two modes on different regions of a camera. Due to the small separation, these two modes were sufficient for a conclusive measurement. In \cite{Yang:16} the measurement was carried out by interference of the emission from two pseudo-thermal sources with a spatially shaped local oscillator (LO), hence implementing a spatial heterodyne detection of the far field. The LO was prepared in either of the two lowest Hermite-Gauss modes; the measured heterodyne current is then proportional to the overlap between the LO and the signal spatial profiles. In \cite{ae2017} instead the authors have implemented the same measurement through a new technique, Super-resolved Position Localisation by Inversion of Coherence along an Edge (SPLICE), which consists in projecting the signal on a mode orthogonal to the initial Gaussian profile, by adding a $\pi$ phase shift to a portion of the beam profile. The displacement between the two sources is imparted via a  Sagnac-like interferometer: a motorized mirror displaces symmetrically both beams, so that the centroid position is kept fixed. A single photon from a heralded source was used. A phase plate is then inserted to project onto the orthogognal mode,  which can be replaced with a slit to operate the classical intensity measurement for comparison. SPLICE does not correspond exactly, however, to projecting in the Hermite-Gauss basis. The scheme proposed in \cite{Tang:16} relies on image inversion interferometry, that is a Mach-Zender interferometric scheme in which image inversion is performed in one of the two arms. The two displaced sources with Gaussian PSFs are sent through the interferometer, and a delay is added on one of the arms to fine-tune the interference. By recording the destructive interference at one of the two outputs it is then possible to infer information on the separation between the two sources. All these implementations have reported immunity to Rayleigh's course: the measured uncertainties on the separation remained well beyond the limit for direct imaging.  

The genuinely multiparameter approach, estimating both the centroid position and the separation at once, has been address in the experiment reported in~\cite{PhysRevLett.121.250503}.
In this experiment, the projection on Hermite-Gauss modes can be replaced by direct imaging, thanks to the adoption of two-photon interference.
Crucially, this amounts to implementing a collective measurement on two copies of the mixed quantum state, which is important to approach the fundamental bound given by the SLD-QFI matrix.
While imperfect visibility in the experiment results in a vanishing precision at zero separation, the scheme is effective in achieving a constant improvement over the precision of direct imaging in a two-parameter protocol.
Further considerations that suggest the adoption of an adaptive scheme have been brought forward in~\cite{Grace2019}, in which the centroid position is treated as a nuisance parameter.

This scheme has inspired extensions to the axial position, explored in~\cite{Zhou:19}, using SLMs to implement a binary mode sorted separating even-order from odd-order Laguerre-Gauss modes, as required by the circular symmetry of the problem. The axial separation, indeed, modifies the width of the PSF. For small axial separations, this measurement, despite its simplicity, grants an uncertainty independent on the measured mean value of the separation. The analogue in the frequency domain has been considered in~\cite{john2018}; there, they consider the frequency or temporal separation between two mutually-incoherent Gaussian pulses. The measurement is performed based on mode-selective upconversion of the signal by means of a quantum pulse gate \cite{Eckstein:11}. As these problems can be mapped exactly to the original imaging problem, the same enhancement is obtained.

This proposal then offers a spectacular advantage with respect to standard techniques; its origin is not properly quantum, instead, it can be traced back to the coherent manipulation at the measurement stage.
However, it suffers a limitation in its fragility with respect to the addition of further control of nuisance parameters.
Imperfections will unavoidably affect the estimation procedure: the unbalance between the intensities of the two sources is among the most relevant in this scenario. A possibility is to keep track of this by including a parameter describing unbalance as a nuisance parameter, instead of conducting a preliminary calibration.
It has been argued that this multiparameter strategy is more indicated to assess the actual working conditions, as well as to to deal with time-dependent conditions~\cite{Vidrighin2014, Crowley2014, Roccia2018}.
The analysis conducted in~\cite{PhysRevA.96.062107} shows that the balanced intensity case is most peculiar, in that no correlations occur between the different parameters. For the more realistic case of unequal intensities, Rayleigh exacts his revenge: the bound on the separation uncertainty diverges as $\lambda_2\rightarrow 0$, nevertheless an advantage can be maintained with respect to the direct imaging approach.
An analytical and numerical investigation, carried out in~\cite{Bonsma_Fisher_2019}, reports how an adapted version of SPLICE can perform well even in the regime of significantly unequal intensities.
This relies on estimating higher order moments of the output probability distribution for a single photon.
This example emphasises how multiparameter treatment helps revealing unsuspected features when including noisy effects.

The extension of the two-source model to an arbitrary number leads to interesting insights.
The works in~\cite{Zhou2019,Tsang2018} have discussed the behaviour of diagonal elements of the SLD-QFI matrix and the possibility of retrieving information by far-field measurements in the subdiffractive regimes.
Quasi-optimal measurement bases have been identified.
A full computation of the SLD-QFI matrix $\mathbf{Q}(\lambdaB)$ for the relative separations $\lambda_\mu=X_{\mu+1}-X_\mu$ of a linear array of $N$ sources has been derived in~\cite{bisketzi2019quantum}, by expanding the operators on a non-orthogonal basis~\cite{Genoni2019}.
It was shown that the $\mathbf{Q}(\lambdaB)$ has rank 2 in the limit of vanishing separations, confirming the previous results of~\cite{Chrostowski2017} obtained by mapping the problem to a qubit model.
In practical terms, this translates into the fact that no more than two parameters can be estimated at once, something reminiscent of Rayleigh's curse.

On the contrary, when the sources are not constrained to be on the same object plane, a multiparameter estimation problem appears naturally, but it is still possible to beat Rayleigh's curse.
This problem was first studied as the estimation of the three-dimensional separation vector of two sources in three-dimensions~\cite{Yu2018}, for a known centroid position and a circular aperture; this makes it suitable to tackle the problem by means of projections onto the basis of Zernicke polynomials~\cite{Zernike345}.
Extensions to the estimation of the full position vectors are found in~\cite{Prasad2019}.

The same three-dimensional problem was studied for arbitrary PSFs in~\cite{Napoli2018}, in the context of surface diagnostics, choosing the transverse and axial coordinates of the two point sources as the relevant parameters.
Denoting $Z_1$ and $Z_2$ as the positions of the sources along the axis of the imaging system, these two parameters are $\lambda_3=(Z_1+Z_2)/2$ and $\lambda_4=(Z_1-Z_2)$.
Also in this case one can expand the operators on a non-orthogonal basis in order to evaluate the SLD-QFI matrix~\cite{Genoni2019}, which, in the limit $\lambda_2\rightarrow 0$ and $\lambda_4 \rightarrow 0$, is diagonal, and also to show that the weak commutativity condition holds true.
This implies that, in this limit, the quantum statistical model is asymptotically classical, i.e. there exists an optimal scheme able to saturate the multiparameter SLD-CRB by means of a collective measurement on an asymptotically large number of copies of the state.

\subsection{Quantum phase imaging}

A transparent object can conveniently be imaged by measuring the optical phase it imparts on the light traversing, it with respect to a reference mode~\cite{Zernike345}. This scheme constitutes the motivation for the works in~\cite{Humphreys2013, Gagatsos2016a, Knott2016, Pezze2017}, in which the simultaneous estimation of multiple phases with quantum light is investigated. This builds over significant effort devoted to the measurement of optical phases with quantum light~\cite{Nagata726}.
The investigation in~\cite{Humphreys2013} considered the estimation of $d$ phases $\lambdaB=\{\lambda_1,\lambda_2, ..., \lambda_d\}$ with a state with a fixed number $N$ of photons. For this class of pure states, the SLDs commute, at least on average on the pure states themselves, hence the model is quasi-classical, saving from the need of collective measurements. 
The treatment can thus be carried out in terms of $\bold{Q}(\lambdaB)$, and in the following we will generally refer to the corresponding matrix SLD-CRB simply as the quantum Cramér-Rao bound (QCRB). One shows that the total variance reaches the optimal value
\begin{equation}
    \Tr[\bold{V}]= C^{\sf S}(\lambdaB,\mathbbm{1}_d) = \frac{(1+\sqrt{d})^2d}{4N^2},
\end{equation}
for the state $\vert\psi_s\rangle =\left(\beta \frac{ (\hat{a}^\dag_0)^N}{\sqrt{N!}}+\alpha\sum_{i=1}^d\frac{(\hat{a}^\dag_i)^N}{\sqrt{N!}} \right)\vert 0\rangle$, where $(\hat{a}^\dag_i)$ is the creation operator on mode $i$ (mode 0 is the reference), $\beta^2=1/(1+\sqrt{d})$, and $\alpha^2=(1-\beta^2)/d$. This state provides an advantage with respect classical light, due to the Heisenberg-like scaling of the total variance as $N^{-2}$. Moreover, it exhibits superior performance with respect to the use of $d$ independent states optimised for single-phase estimation, keeping the total number of photons fixed; in that case the total variance would assume the value $d^3/N^2$. 

Finding measurements achieving the QCRB is a non-trivial task: an example of one such measurement was reported in~\cite{Humphreys2013}, ensuring it could provide separate values for each $\lambda_\mu$. Notably, this requires a projector back on the original state; the most general conditions for the optimal measurement have been obtained and discussed in~\cite{Pezze2017}.

The same problem has also been investigated in~\cite{Gagatsos2016a} considering Gaussian input states with quadrature squeezing. These are then superposed in a linear interferometer to achieve the optimal possible state. There it was shown that the Fisher information for simultaneous estimation of the phases can only provide a constant advantage by a factor 2. This has been attributed to a limitation intrinsic to the Gaussian form of the inputs. Interestingly, Ref.~\cite{Knott2016} has succeeded in showing that, if the phase shifts are reparametrised as phase differences between consecutive arms, the same precision can be attained by means of local as well as global estimation strategies.

An integrated realisation of multiphase estimation has been tackled in~\cite{Polino2019}. There, they perform phase estimation in a reconfigurable integrated three-modes interferometer, with a two-photon input state. The device consists of two tritters; these are optical elements performing linear coupling between two sets of three modes~\cite{chiara13}.
Phase shifts can be added on each of the three modes independently, in between the two tritters.
The implemented scheme follows closely the original formulation in~\cite{Humphreys2013}, consisting in one mode acting as a reference, while the parameters to be estimated are encoded in the two phase differences with respect to the other two modes.
The two-photon input state is prepared with the first tritter, operating a unitary matrix $U_{\tiny A}$, which can be tailored using the tritter's thermo-optic phase; then the state is propagated through the three modes, where thermo-optic phases are set to configure the two parameters to be estimated.
Eventually the state is measured by undergoing the second tritter, performing the inverse unitary transformation $U_{\tiny A}^\dag$, such that $U^\dag_{\tiny A} U_{\tiny A}=\mathbbm{I}$.
Coincidence counting is then performed between the outputs of the circuit.
In a general interferometer, this may not correspond to the optimal measurement, but it shows practical advantages in terms of ease of fabrication and manipulation~\cite{Pezze2017}.
The  detailed theoretical analysis of the system capabilities has been extensively carried out in~\cite{Ciampini2016,spagnolorep}.
The parameters are then estimated with a maximum likelihood routine, and the result show good agreement with the theoretical model, validating the suitability of the chosen platform for the multiparameter estimation protocol.


\subsection{Discussion and perspectives}
Quantum sensing has witnessed continuous progress over the last decades. The focus on single-parameter estimation has allowed to build a large toolkit of experimental and theoretical methods, however this limits the class of systems which can be investigated.  We have reviewed a particular application, quantum imaging, falling intrinsically fall in the multiparameter domain. Along this, there exist many others connected to quantum network diagnostics, spectrally resolved optical phase
profile, and magnetic field mapping in time and space.

Access to the novel applications of multiparameter estimation demands for a deep rethinking of the methods that underlie the design of standard quantum sensors. It will become crucial to technological advancements the merging of theoretical insights and experimental best practices into a series of design criteria of practical relevance. This must be informed on the specific needs and peculiarities of the sought application. Establishing a common framework with potential end-users, including biologists, atmospheric physicists, material scientist and the like, will be a pressing urge in the forthcoming years. 

To illustrate our point we consider the example of simultaneous estimation of axial and transverse separation of two sources~\cite{Napoli2018}. 
In this context, the work in~\cite{Lupo2019} has made a bridge between superresolution imaging and linear interferometry, 
by showing that, in the paraxial approximation and for particular configurations of the collectors, a simple linear interferometer with photo-detection is optimal to estimate both parameters (also assuming that the position of the centroid is known).
However, the effectiveness of this approach for the general multiparameter problem and for more sources has still to be fully investigated.
As a matter of fact the main conceptual challenge consists in finding the measurement strategy that exhibits simultaneous advantage for both parameters, while remaining experimentally feasible.
This should prove to be more effective than simply alternating between the two optimal basis for individual parameter estimation and it is a central issue for establishing the advantage of multiparameter quantum sensing, studied in several contexts~\cite{Vaneph2012,Humphreys2013,Gagatsos2016a,Baumgratz2015,Ragy2016,Nichols2018,Kura2017,Gorecki2019,Chen2019}.

Further, the possibility of introducing nuisance parameters should be taken into account; in different contexts it has been shown how multiparameter estimation is is significantly affected by the introduction of extra noisy parameters~\cite{Roccia2017,PhysRevLett.121.250503,PhysRevA.96.062107}.
An analysis in this sense is yet to be conducted. 
Broadening the vision from this particular instance, we have two lessons to learn: the first is that in multiparameter estimation the lack of a straightforward prescription based on the SLD opreators is a drawback; so far, the most successful approaches have elaborated measurement strategies on the basis of experimental convenience and then used the QFI treatment as a valdiation method.
The second aspect concerns extending the capabilities of the schemes to encompass additional control.
Naively, monitoring additional nuisance parameters may improve the stability of the estimation, a task which is relevant for moving towards real-life devices.
In this respect, one should also consider the performance of sensors based on time-continous monitoring; the corresponding mathematical framework needed to assess their optimality in terms of multi-parameter quantum estimation has already been established~\cite{GammelmarkPRA,GammelmarkPRL}, and single-parameter theoretical results have been recently obtained in different contexts~\cite{KiilerichPRA,AlbarelliNJP,AlbarelliQuantum}.
We also remark that for continuous monitoring the classical data processing involved in estimation is highly non-trivial, see e.g.~\cite{Ralph2017}.
We expect that these tools will turn to be useful to assess the performance of multi-parameter time-continuous sensors and for the characterization of devices, such as superconducting circuits useful for quantum technologies~\cite{Ficheux2017}. 
On a more sophisticated level, this kind of multiparameter approach is still to be fully integrated within a proper open loop or closed loop control scheme for improving the performance of the estimation actively, as already suggested in~\cite{Yuan2016b,Liu2017d}.
Further insight could possibly be gained by casting quantum imaging as a problem in distributed sensing~\cite{Proctor2018,Eldredge2018,Ge2017,Zhuang2018c,guo2019}.

Finally, when approaching real-life scenarios, it may be required to explore non-asymptotic regimes due to limited amount of resources available. In these instances, Bayesian adaptive protocols are useful to achieve optimality, as already accomplished for single parameter estimation~\cite{Lumino,paesani17,rubio2019limited}.
Recent theoretical investigations suggest this may also be the case when addressing the multiparameter case~\cite{Rubio2019}.
Interestingly, for optimal single parameter metrological protocols the discrepancy between a Bayesian analysis and the QCRB does not disappear even in the asymptotic limit~\cite{Gorecki2019a} and one should expect the same behaviour also in the multiparameter case.

As a concluding remark, we are persuaded that the potential of multiparameter estimation for quantum technologies has barely been hinted in the applications performed to date.
We anticipate in the next years it will deliver innovative results on both fundamental issues and technological applications, more deeply rooted into theoretical advancements.

\section*{Acknowledgements}
We are grateful to G.~Adesso, E.~Bisketzi, D.~Branford, A.~Datta, J.~F.~Friel, A.~Fujiwara, S.~Moreno, M.~G.~A.~Paris, L.~Pezzè, L.~L.~Sanchez-Soto, F.~Sciarrino, N.~Spagnolo, {\AE}.~Steinberg, J.~Suzuki, N.~Treps, M.~Tsang, T.~Tufarelli, J.-P.~Wolf and J.~Yang for fruitful discussion.
FA acknowledges support from the UK National Quantum Technologies Programme (EP/M013243/1) and from the National Science Center (Poland) grant No. 2016/22/E/ST2/00559.
MGG acknowledges support from a Rita Levi-Montalcini fellowship of MIUR.


\nocite{apsrev41Control}
\bibliography{persbiblio}

\end{document}